# Reflectivity spectra as absorption ones – is it correct ?

M.A. Andreeva, R.A. Baulin

*Faculty of Physics, Lomonosov Moscow State University, 119991, Moscow, Russia*

**Synopsis** - Linear dependence of the reflectivity spectra on the resonant absorption coefficient are obtained in the total external reflection region but only for ideal cases of low-absorptive homogeneous resonant media or ultrathin resonant layers. Restrictions for simplified treatment of the reflectivity spectra as absorption ones are revealed by theoretical analysis and model calculations.

**Abstract** - Approximate expressions for X-ray resonant and Mössbauer reflectivity in the total external reflection region are developed for the limiting cases of a semiinfinite mirror with a small resonant addition to the total susceptibility and for the case of an ultrathin resonant layer. It is shown that in this region the reflectivity can really show the linear dependence on the imaginary part of resonant susceptibility, therefore in these cases the consideration of reflectivity spectra (R-spectra) as absorption resonant spectra, undertaken in several experimental works, can be justified. However, we have selected several effects producing essential dispersive distortions of the R-spectrum shape even for very small grazing angles. It has been shown that the dispersive corrections to the R-spectrum shape are mostly essential if the nonresonant absorption is relatively large. Model calculations demonstrate that the quantitative spectroscopic information extracted from R-spectra by using the software developed for absorption spectra can be inaccurate.

1.      Introduction

X-ray spectral diagnostic (XANES, XMCD, XNCD, XRMR, DAFS, as well as Mössbauer spectroscopy) actively developing at synchrotrons opens a new era in investigations of chemical bonds, electronic structure of atoms, element specific magnetic properties for samples in different environments (temperature, pressure, magnetic and electric fields). Absorption spectra (measured directly or by fluorescence or electron total yield) can be straight used for identification of different transitions between electron shells, for application of sum rules determining spin and orbital contributions to the atomic magnetic moment, for extracting hyperfine interaction parameters in the case of Mössbauer spectra and so on. There are a lot of computer procedure and software packages for treating absorption spectra (like sum rules (Thole *et al*., 1992; Carra *et al*., 1992; O'Brien & Tonnere, 1994), Mössbauer UNIVEM-*MS* program and MossWinn code (Klencsár *et al*., 1996), CONFIT2000 (Žák & Jirásková, 2006), SpectrRelax (Matsnev & Rusakov, 2012) and so on).

In many cases, e.g. for investigations of thin resonant layers or surfaces with synchrotron radiation (mostly suitable for angular resolved measurements), the spectra are measured in reflection geometry at small grazing angles. Contrary to the absorption spectra depended only from the imaginary part of the refractive index, reflectivity spectra (R-spectra) depend on the real and imaginary parts of the reflective index in a quite complicated manner through Fresnel formula (Born & Wolf, 1968), Parratt algorithm (Parratt, 1954), or 4×4 matrix formalism for anisotropic multilayers (Borzdov *et al*., 1976; Azzam & N. Bashara, 1977; Zak *et al*., 1991; Irkaev *et al*., 1994; Röhlsberger, 1999; Stepanov & Sinha, 2000). Therefore, the extracting spectroscopic information from R-spectra is not direct and is connected with a time-consuming fit procedure. For relatively large glancing angles ($\theta > 2 \div 3\ \theta_c$, $\theta_c$ is the critical angle of the total external reflection) the kinematic approach of the X-ray reflectivity theory (Hamley & Pedersen, 1994) can be applied which allows simpler treating R-spectra (for Mössbauer Bragg reflections from periodic multilayers it had been analyzed in e.g. (Andreeva & Lindgren, 2005)). In the total reflection region the simple kinematic approximation cannot be applied, at the same time in this region R-spectra look like absorption spectra (specific shapes of R-spectra for different grazing angles were investigated long ago (Bernstein & Campbell, 1962; Kao *et al*., 1994)), and in many papers they are treated by software developed for absorption spectra fit (see e. g. (Mitsui *et al*., 2012; Mitsui *et al*., 2016; Cini *et al*., 2018; Cucinotta *et al*., 2020; Mitsui *et al*., 2020; Cini *et al*., 2020; Okabayashi *et al*., 2021a; Yokota *et al*., 2021; Okabayashi *et al*., 2021b)). However, no mathematical support for such procedure has been developed for the region of the total external reflection.

Here we thoroughly analyze the R-spectrum shape in the total reflection region for X-ray radiation and outline the limits of the applicability of the simplest procedure for its fit based on the absorption software.

## 2. Semiinfinite mirror

### 2.1. Total reflection region

For the case of isotropic mirror and grazing angles of incidence $\theta$ the Fresnel formula for the reflectivity amplitude is independent on the polarization of radiation and is given by the well-known expression (Born & Wolf, 1968):

$$R^{Fr} = \frac{\sin\theta - \sqrt{\sin^2\theta + \chi}}{\sin\theta + \sqrt{\sin^2\theta + \chi}}, \qquad (1)$$

where $\chi$ is a scalar susceptibility which is supposed to be constructed from nonresonant ($^{nr}$) and resonant ($^{res}$) parts:

$$\begin{aligned}\chi &= \chi^{nr} + \chi^{res}(\omega); \\ \chi^{nr} &= -2\delta^{nr} + 2i\beta^{nr}, \quad \chi^{res}(\omega) = -2\delta^{res}(\omega) + 2i\beta^{res}(\omega)\end{aligned} \qquad (2)$$

Here we use the symbols $\delta$ and $\beta$ which are the conventional designations of the corrections to the refractive index $n = 1 - \delta + i\beta = 1 + \chi/2$ for X-ray radiation, $\omega$ is the photon frequency, for simplicity here it is supposed that nonresonant part $\chi^{nr}$ does not depend on the photon energy in the narrow energy interval of resonant transitions (for Mössbauer spectra it is always true).

We suppose that in the region of the total reflection $\operatorname{Re}\chi^{nr}$ is large enough compared with $\sin^2\theta$ and negative, so the designation is inserted:

$$\sqrt{\sin^2\theta - 2\delta^{nr}} = i\rho, \qquad (3)$$

and the following approximated expression for the square root is obtained:

$$\sqrt{\sin^2\theta + \chi} \cong i\rho + i\delta^{res}/\rho + \beta/\rho. \qquad (4)$$

It has been supposed that $\beta = \beta^{nr} + \beta^{res} \ll \rho^2 = 2\delta^{nr} - \sin^2\theta$ and $\delta^{res} \ll \rho^2$. With approximation (4) the reflected intensity can be presented by the following expression (the detailed calculations for this and for the following expressions are given in the Appendix):

$$\left|R^{Fr}(\omega)\right|^2 \cong 1 - \frac{4\sin\theta}{\rho\left|\operatorname{Re}\chi^{nr}\right|}(\beta^{nr} + \beta^{res}(\omega)). \qquad (5)$$

It is nice to mark that the obtained expression really establishes the linear dependence of the reflected intensity on the imaginary part $\beta^{res}$ of the resonant part of the total susceptibility. Therefore, this expression legalizes the attempts to interpret the R-spectra in the total reflection region as the absorption spectra and justify the usage the common software for their fit, undertaken in e.g. (Mitsui *et al*., 2012; Mitsui *et al*., 2016; Cini *et al*., 2018; Cucinotta *et al*., 2020; Mitsui *et al*., 2020; Cini *et al*., 2020; Okabayashi *et al*., 2021a; Yokota *et al*., 2021; Okabayashi *et al*., 2021b).

However, several effects disturb the simplest linear connection between the reflectivity and absorption for the angles in the total reflection region given by the expression (5).

*The first effect is the influence of a nonresonant absorption.* If $\beta^{nr} \gg \beta^{res}$ the expression (5) transforms to a more complicated form:

$$\left|R^{Fr}(\omega)\right|^2 \cong \frac{(\sin\theta - \beta^{nr}/\rho)^2 + \rho^2}{(\sin\theta + \beta^{nr}/\rho)^2 + \rho^2} - \beta^{res}(\omega)\frac{4\sin\theta(2\delta^{nr} - (\beta^{nr}/\rho)^2)}{\rho\left[(\sin\theta + \beta^{nr}/\rho)^2 + \rho^2\right]^2} +$$
$$+ \delta^{res}(\omega)\frac{8\sin\theta\,\beta^{nr}}{\rho\left[(\sin\theta + \beta^{nr}/\rho)^2 + \rho^2\right]^2}. \qquad (6)$$

The expression (6) includes not only a change in the nonresonant background (first term), but also a contribution from the real part of the resonant susceptibility $\delta^{res}$ to the R-spectrum (third term in (6)), adding a dispersion correction to the R-spectrum shape. It is important to note that this dispersion term is proportional to the nonresonant absorption $\beta^{nr}$.

For an artificial case $\beta^{nr} = 0$ the R-spectra calculated for grazing angles in the total reflection region by the exact formula (1) and approximate expressions (5) and (6) are practically identical and their shape reproduces the energy dependence of $\beta^{res}$ (i.e. the shape of an absorption spectra). However, even a relatively small nonresonant absorption ($\beta^{nr} \neq 0$) introduces a dispersion distortion of the R-spectrum shape (Fig. 1a), and the further artificial increase of $\beta^{nr}$ makes the asymmetry of the resonant line in R-spectra more clear, besides the shift from the exact resonance appears (see Fig. 1b). For this case the difference between the exact calculations and calculations by the approximate expressions (5) and even (6) becomes essential.

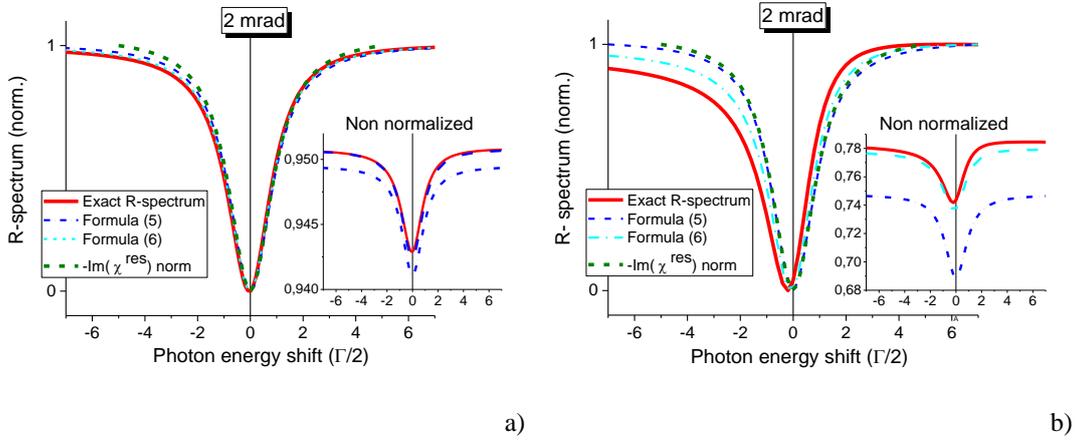

a) b)

**Figure 1**

*R-spectra for $\theta = 2$ mrad in comparison with the shape of the absorption spectra, calculated for the simplest resonant addition $\chi^{res} = \dfrac{-0.1}{x+i} 10^{-6}$ to the nonresonant susceptibility $\chi^{nr} = (-14.6 + i\, 0.6) \times 10^{-6}$ (this $\chi^{nr}$ approximately corresponds to an iron for Mössbauer wavelength $\lambda=0.086$ nm) (a) and when the nonresonant absorption is increased in 5 times ($\chi^{nr} = (-14.6 + i\, 3.0) \times 10^{-6}$) (b).*

*The second effect is the line interference.* Probably the asymmetric distortion of the resonant line is not so important for the spectroscopy based on the reflectivity data, but this distortion leads to the defacement of the relative weight of different contributions. In absorption spectra different contributions are independent from each other, in R-spectra the overlap of the lines leads to the mutual influence. That is demonstrated by the picture with two resonance lines of equal amplitude: calculated R-spectrum in Fig. 2 shows different intensity for two lines. Note, that this effect has been demonstrated in the first paper devoted to the Mössbauer reflectivity (Bernstein & Campbell, 1962). Therefore, the treatment of the complicated R-spectra with the programs developed for the absorption spectroscopy will give the wrong weights of contributions.

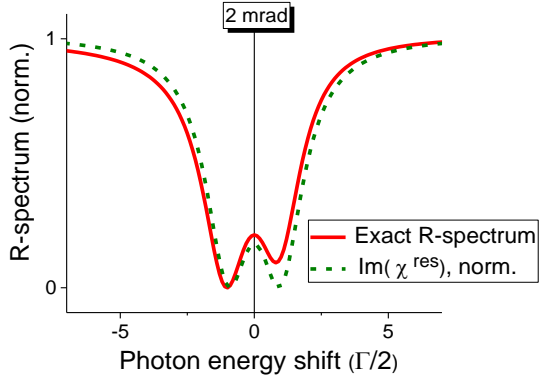

**Figure 2**

*R-spectra, calculated for* $\chi^{res} = -1.0 \times (\frac{1}{x+1+i} + \frac{1}{x-1+i}) \times 10^{-6}$ *and* $\chi^{nr} = (-14.6 + i\,0.6) \times 10^{-6}$ *in comparison with the doublet shape of the absorption spectra.*

*The third effect is the sequence in the depth position of different contributions.* It is well-known that the very small penetration depth in the total reflection region changes the relative weight of contributions from resonant atoms (nuclei) placed at different depths in a multilayer. This effect is illustrated by Fig. 3a. This depth selectivity of the reflectivity method is well known and has been effectively used in numerous papers (see e.g. (Bernstein & Campbell, 1962; G. Martens & P. Rabe, 1980; Andreeva *et al*., 1991; Irkaev *et al*., 1994; Freeland *et al*., 1997; Geissler *et al*., 2001; Andreeva *et al*., 2006; Andreeva *et al*., 2015)).

*The fourth effect is the dichroic scattering.* The anisotropy of the resonant scattering supposes that the scattering wave can have another polarization than that in the incident wave. In this case the reflectivity is described by more complicated formalism than the Fresnel formula (the theory of X-ray reflectivity from anisotropic multilayers is presented in e.g. (Borzdov *et al*., 1976; Azzam & N. Bashara, 1977; Zak *et al*., 1991; Irkaev *et al*., 1994; Röhlsberger, 1999; Stepanov & Sinha, 2000)). The contribution of the rotated polarization to the reflected signal changes the ratio of lines in Mössbauer R-spectra. This effect can not be taken into account in the single absorption process and consequently the treatment of the R-spectra as absorption spectrum gives the wrong information about the magnetic field orientation (Fig. 3b). Calculation of the R-spectra is done for the $B_{hf}$ azimuth angle $\gamma \cong 30°$ relative the beam direction, but the ratio of the 1st and 2nd line intensities will correspond $\gamma \cong 40°$ if the fit of this spectra will be done by the program for the absorption spectrum treatment.

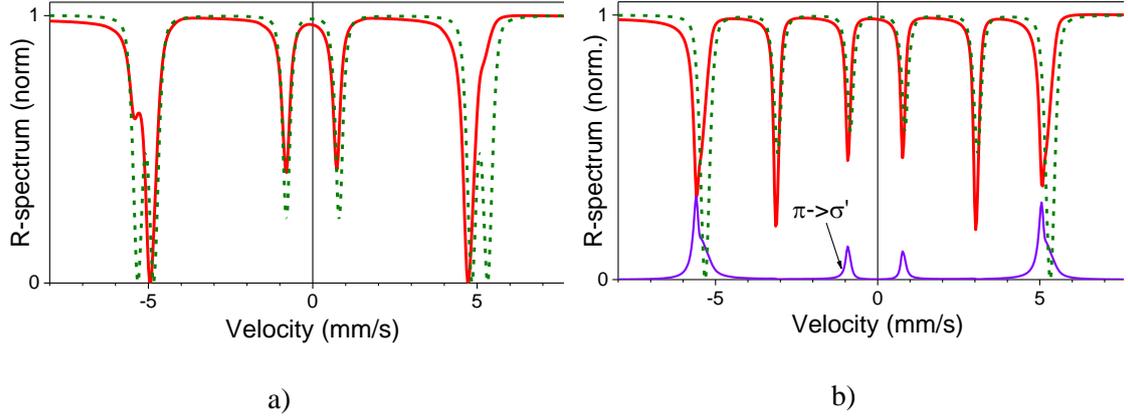

**Figure 3**

*Comparison of the Mössbauer R-spectra (thick solid line) and the absorption one (dashed line), calculated (a) for the two-layer $^{57}Fe$ film (50 % enrichment): 5 nm $B_{hf}$=30 T + 5 nm $B_{hf}$=33 T, on a glass substrate (solid line), magnetic hyperfine fields $B_{hf}$ in both sublayers are oriented perpendicular to the surface and (b) for the 10 nm Fe film ($^{57}Fe$ enrichment 50%, $B_{hf}$=33 T) magnetized in the surface plane with the azimuth angle 30º relative to the beam direction. Thin line in (b) is the contribution of the reflectivity with rotated π→σ′ polarization to the R-spectrum. Both R-spectra calculated for θ = 2 mrad, for π-polarized incident radiation. Calculations are done by our REFSPC software (Andreeva, 2008; http://www.esrf.eu/Instrumentation/software/data-analysis/OurSoftware/REFTIM-1).*

## 2.2. Exact Critical angle

For completeness we calculate the approximate expression for the Fresnel reflectivity at the exact critical angle when

$$\sin^2\theta - 2\delta^{nr} = 0. \qquad (7)$$

For small resonant addition $\delta^{res}(\omega)$ and $\beta^{res}(\omega)$ to the refraction index the reflectivity is described by the formular

$$\left|R^{cr}(\omega)\right|^2 \cong \left|R^{nr}\right|^2 + \frac{\sqrt{2\delta^{nr}\beta^{nr}}}{(\delta^{nr}+\beta^{nr}+\sqrt{2\delta^{nr}\beta^{nr}})^2}[\delta^{res}(\omega)(1+\delta^{nr}/\beta^{nr}) + \beta^{res}(\omega)(1-\delta^{nr}/\beta^{nr})], \qquad (8)$$

where

$$\left|R^{nr}\right|^2 \cong \frac{\delta^{nr}+\beta^{nr}-\sqrt{2\delta^{nr}\beta^{nr}}}{\delta^{nr}+\beta^{nr}+\sqrt{2\delta^{nr}\beta^{nr}}}. \qquad (9)$$

Supposing that for the right and left circular polarization, being the eigen polarization in the L-MOKE geometry, the resonant magnetic addition to the nonresonant susceptibilities has different signs:

$$\chi^{res}_{\pm}(\omega) = \pm(-2\delta^{res}(\omega) + 2i\beta^{res}(\omega)), \qquad (10)$$

then the asymmetry in the reflectivity $A(\omega)$ at the exact critical angle can be presented by the expression:

$$A(\omega) = \frac{|R^+|^2 - |R^-|^2}{|R^+|^2 + |R^-|^2} \cong \frac{\sqrt{2\delta^{nr}\beta^{nr}}}{\delta^{nr2} + \beta^{nr2}} [\delta^{res}(\omega)\left(1 + \frac{\delta^{nr}}{\beta^{nr}}\right) + \beta^{res}(\omega)\left(1 - \frac{\delta^{nr}}{\beta^{nr}}\right)], \quad (11)$$

which is almost the same as was given in the paper (Neumann *et al.*, 1998) (with a slight correction negligible when $\beta^{nr} \ll \delta^{nr}$). A popular opinion is that at the exact critical angle the R-spectrum presents the dispersive part $\delta^{res}(\omega)$ of the refraction index. As follows from (8), (11) that is not true: the R-spectrum shape and the polarization asymmetry is mainly determined by $\delta^{res}(\omega)$ only when $\delta^{nr}/\beta^{nr} \sim 1$, otherwise both the absorption and dispersive parts of the resonant addition to the refractive index ($\delta^{res}(\omega)$ and $\beta^{res}(\omega)$) influence this spectral dependence at the critical angle.

If $\beta^{nr} \ll \delta^{nr}$ with notation $\beta = \beta^{nr} + \beta^{res}$ the reflectivity at the exact critical angle takes a form:

$$|R^{cr}(\omega)|^2 \cong 1 - \frac{2\sqrt{2}\sqrt{\sqrt{\delta^{res}(\omega)^2 + \beta(\omega)^2} - \delta^{res}(\omega)}}{\sqrt{\delta^{nr}}} . \quad (12)$$

For $\beta = 0$ and $\delta^{res} > 0$ from (12) it follows $|R^{cr}|^2 = 1$, but if $\delta^{res} < 0$ then $|R^{cr}|^2 \cong (1 - 4\sqrt{\delta^{res}/\delta^{nr}})$.

3. **Ultrathin resonant layers**

In several papers the resonant spectra were measured from ultrathin resonant layers. It had been shown that the reflectivity amplitude from an ultrathin resonant layer placed under a reflecting medium can be described by the formula (Andreeva & Lindgren, 2002; Andreeva *et al.*, 2019):

$$R^{tot}(\omega) \cong R^s + E^2 r^d(\omega) , \quad (13)$$

where $R^s$ is the reflectivity amplitude from an underlying substrate, $E = 1 + R^s$ is the total electric field at the position of the resonant layer (if we take the amplitude of the incident wave equals 1) and $r^d(\omega)$ is the reflectivity amplitude of the ultrathin resonant layer in vacuum described by the well-known expression:

$$r^d(\omega) = \frac{i\pi d}{\lambda \sin \theta} \chi^d(\omega), \quad (14)$$

where $\chi^d = -2\delta^d + 2i\beta^d$ is the susceptibility of an ultrathin layer having thickness $d$, $\lambda$ is the x-rays wavelength.

If the frequency dependence of $r^d(\omega)$ in (14) is presented by a pure one resonance case $r^d(\omega) \propto \dfrac{-i}{x+i}$, where $x = \dfrac{\hbar(\omega - \omega_0)}{\Gamma/2}$, $\Gamma$ is the full line-width, $\omega_0$ is the exact resonant frequency, then the reflectivity from the ultrathin layer without any support is given by $|r^d|^2 \propto \dfrac{1}{x^2+1}$ which corresponds to the emission line but not to the absorption one. If some underlying substrate exists the shape of the resonant line in the R-spectrum becomes different.

If $r^d(\omega)$ is small enough then the total reflectivity $|R^{tot}(\omega)|^2$ can be described by the following expression:

$$|R^{tot}(\omega)|^2 = (R^S + E^2 r)(R^{S*} + E^{*2} r^*) \cong$$
$$\cong |R^S|^2 + 2[R' + (2+R')|R^S|^2] r' + 2R''(1 - |R^S|^2) r'' = \quad , \qquad (15)$$
$$= |R^S|^2 + Q_1 r'(\omega) + Q_2 r''(\omega)$$

where we use ' and " for selection of the real and imaginary parts of $R^S = R' + iR''$ and $r^d(\omega) = r' + i r''$. The notations $Q_1$ and $Q_2$ are used in (15) for the nonresonant coefficients determining the contribution of the real and imaginary parts of $r(\omega)$ to the total reflectivity $|R^{tot}(\omega)|^2$. Note that due to the presence of the imaginary unit $i$ in the expression for $r^d$ (14), that is the real part of $r^d$ (with coefficient $Q_1$) determines an absorption-like dependence of $|R^{tot}(\omega)|^2$, e.g. for a single resonance $r' \propto \mathrm{Re}\dfrac{-i}{x+i} = \dfrac{-1}{x^2+1}$, and the relative value of $Q_2$ determines the dispersion-like corrections to the resonant R-spectrum shape $|R^{tot}(\omega)|^2$. From the general view of (15) it is clear that in the total reflection region where $|R^S|^2 \approx 1$ the contribution of the dispersive part to the total reflectivity $|R^{tot}(\omega)|^2$ is almost negligible ($\propto Q_2 \propto (1 - |R^S|^2) \approx 0$). It can have some influence on $|R^{tot}|^2$ only for a strongly absorptive substrate when $|R^S|^2 < 1$ in the total reflection region.

If $R^S$ in (15) is the Fresnel reflectivity amplitude from substrate, for separation of its real and imaginary parts the formula (1) can be rewritten in the form:

$$R^S = R' + iR'' = \dfrac{1}{W}\left((1 - \sqrt{u^2 + v^2}) - i\sqrt{2}\sqrt{\sqrt{u^2+v^2} - u}\right), \qquad (16)$$

where

$$u = 1 - 2\delta^s / \sin^2\theta, \quad v = 2\beta^s / \sin^2\theta,$$
$$W = 1 + \sqrt{u^2 + v^2} + \sqrt{2}\sqrt{\sqrt{u^2 + v^2} + u} \qquad (17)$$

($\delta^s$ and $\beta^s$ characterize the susceptibility of the substrate $\chi^s = -2\delta^s + 2i\beta^s$). For $Q_1$ and $Q_2$ in this case we get (see the Appendix):

$$Q_1 = 2\left(R' + (2 + R')|R^s|^2\right) = \frac{8(1-u)}{W^2} = \frac{16\delta^s}{W^2 \sin^2\theta}, \qquad (18)$$

$$Q_2 = 2R''\left(1 - |R^s|^2\right) = -\frac{8}{W^2}v = -\frac{16\beta^s}{W^2 \sin^2\theta}. \qquad (19)$$

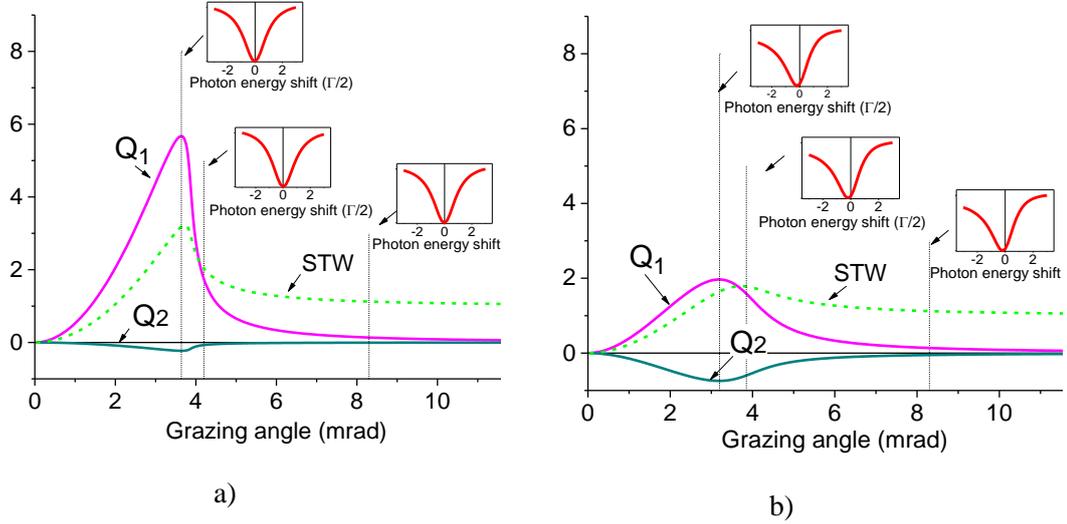

**Figure 4**

*Angular dependence of the $Q_1$ and $Q_2$ coefficients determining the absorptive and dispersive parts forming the reflectivity spectra from ultrathin resonant layer placed on a nonresonant substrate. The shape of the single resonance line in normalized R-spectra is drawn for selected grazing angles (marked by the thin vertical lines) in the inserts. Dashed line shows the square module of the radiation total electric field $|E|^2$, i.e. standing wave (STW), on the substrate surface. Calculations for $\chi^d = (-5.0 + i\, 0.01) \times 10^{-6} + \frac{-0.1}{x+i} \times 10^{-6}$, d = 0.001 nm, $\chi^s = (-14.8 + i\, 0.6) \times 10^{-6}$ in (a) and $\chi^s = (-14.6 + i\, 5.6) \times 10^{-6}$ in (b); λ=0.086 nm.*

It is surprising that an absorption-like part of $|R^{tot}(\omega)|^2$ (~$Q_1 r'$) is proportional to the real part of the substrate susceptibility $\delta^s$ and the dispersive term in $|R^{tot}(\omega)|^2$ (~ $Q_2 r''$) is proportional to the imaginary part of the substrate susceptibility $\beta^s$. So, if the substrate is nonabsorptive, the R-spectrum will have a pure linear dependence on $\beta^d$. In another words, in this case the R-spectrum of resonant ultrathin layer can be interpreted as the absorption spectrum. The dispersion distortion of the R-spectrum depends on the absorption value in a nonresonant substrate.

Even more surprising is that

$$Q_2/Q_1 = const = -\beta^s/\delta^s, \qquad (20)$$

i.e. the ratio of the absorption and dispersion terms in $\left|R^{tot}(\omega)\right|^2$ does not depend on the angle of incidence (Fig. 4a). Therefore, the shape of R-spectrum from ultrathin layer placed on the substrate does not depend on the angle of incidence and its asymmetry is determined only by this ratio $\beta^s/\delta^s$. Thai is probably the reason why in some papers, in which Mössbauer R-spectra have been measured for ultrathin resonant layers, the angles for incident radiation has not been pointed out. For relatively large absorption in substrate R-spectra acquire some dispersion addition. That is illustrated by Fig. 4b.

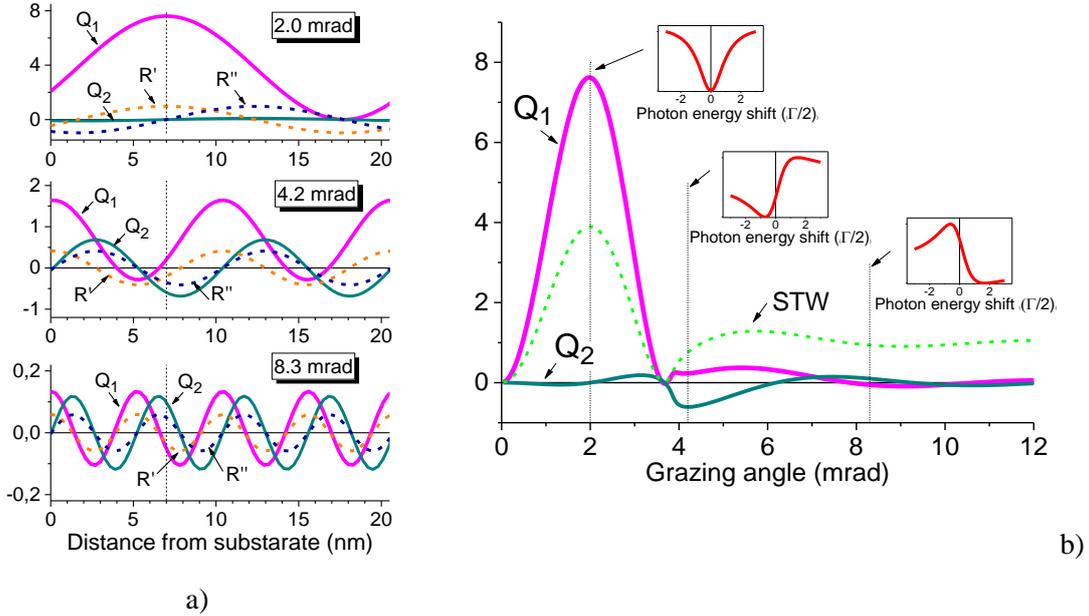

**Figure 5**

*(a) Variations of $R'$ and $R''$, determined by (21), and $Q_1$ and $Q_2$, calculated by general expression (15), with increasing distance H of ultrathin layer from substrate for selected grazing angles. (b) Angular dependence of the $Q_1$ and $Q_2$ at the distance from substrate H=7 nm (marked in (a) by thin vertical line). The shape of the single resonance line in R-spectra from ultrathin resonant layer, placed at distance 7 nm from substrate is drawn for selected grazing angles (2 mrad, 4.2 mrad and 8.3 mrad) is presented in the inserts. Calculations for the same $\chi^d$ as in Fig. 4 and $\chi^s = (-14.6 + i\, 0.6) \times 10^{-6}$.*

The illustrations in Fig. 4 refer only to the case when an ultrathin layer is placed directly on a substrate, giving Fresnel reflectivity. If the layer is placed on some distance $H$ from a substrate the relation (20) is not true. (Such situation can be approximately fulfilled for molecules used in Langmuir-Blodgett method having resonant atoms in their head, like it was considered in (Bedzyk et al., 1989), or for e.g. low density small particles covered by a resonant isotope (Merkel et al., 2015)). In this case $R^s$ in (15) acquires an additional phase shift:

$$R^s \Rightarrow R = R^s e^{iqH} = R' + iR'', \qquad (21)$$

where $q = \dfrac{4\pi}{\lambda}\sin\theta$ is the scattering vector. According to (21) $R'$ and $R''$ oscillate with the increasing of the distance $H$ with a shift of the phase (Fig. 5a), consequently for $Q_1$ and $Q_2$ the more complicated expressions take place. The shape of R-spectrum follows the variations of $Q_1$ and $Q_2$ with the change of the angle $\theta$ (Fig. 5b): the dispersion addition to the shape of R-spectrum becomes more pronounced when $|Q_2| > |Q_1|$. However, in the total reflection region always $Q_1 \gg Q_2$ and R-spectra looks like the absorption spectra.

## 4. Example of experimental data treatment

In the presented analysis for clarity the different factors disturbing the simple absorption-like shape of R-spectra in the total reflection region are considered individually while for spectroscopy of the real multilayer samples all of them can influence on the experimental R-spectra and insert errors into the result of their interpretation based on the pure absorption concept.

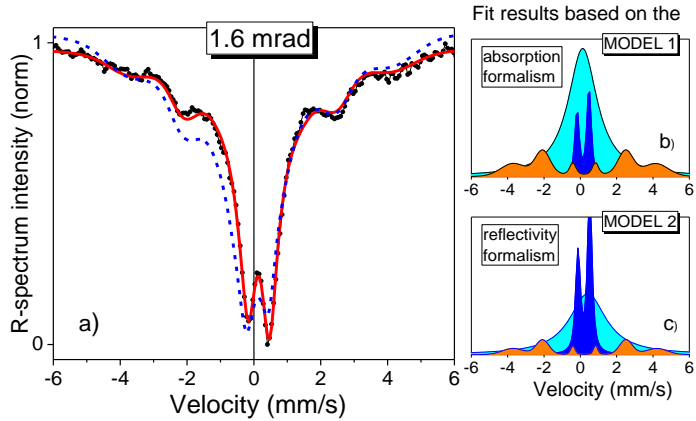

**Figure 6**

*(a) Experimental R-spectrum (Nosov et al., 2022) (symbols) measured at θ=1.6 mrad for a thin film $YFeO_3$ (≅ 10 nm, 35% $^{57}Fe$ enrichment) on $Al_2O_3$ substrate. Fit results based on the absorption and reflectivity algorithm for its description give a set of hyperfine sub-spectra with different weights ((b) and (c) respectively) though both equally satisfy the experimental R-spectra (solid line in (a)). The theoretical spectra calculated by the exact reflectivity theory with program REFSPC (Andreeva, 2008) for the model 1 is presented in (a) by dashed line.*

An example of the noncorrect experimental data analysis is shown in Fig. 6. Experimental R-spectrum were measured at ID18 of ESRF by our colleagues from IMP UB RAS (Proposal HC-4300) (Nosov *et al.*, 2022).

**5.      Conclusion**

In summary, we have derived the approximate analytical expressions for the X-ray resonant reflectivity in the total reflection angular region which validate the application of the software developed for the absorption spectra to the R-spectrum fit in some ideal cases (very small resonant contribution to the refractive index, ultrathin resonant layers). It is confirmed that in the case of ultrathin resonant layers the R-spectrum shape is really corresponded to the absorption spectrum for the angles in the total external reflection region. At the same time the requirement of a homogeneous low absorptive resonant media or of an ultrathin thickness of the resonant layer is hardly fulfilled for real samples. The restrictions for absorption approach to R-spectra are thoroughly analyzed. Finally, it has been demonstrated that the quantitative spectroscopic analysis of R-spectra should be done on the basis of the correct reflectivity theory.

**Acknowledgements**

The authors would like to thank Prof. A.P. Nosov (Mikheev Institute of Metal Physics of the Ural Branch of RAS, Ekaterinburg, Russia) for granting the experimental data which fit stimulates this work.

**Appendix**

1. *Reflectivity from a low absorption semi-infinite medium in the presence of a small resonant scattering in the total external reflection region*

For semi-infinite mirror with scalar susceptibility $\chi$

$$\chi = \chi^{nr} + \chi^{res}(\omega) , \qquad (S1)$$

where the resonant part $\chi^{res}(\omega) = -2\delta^{res}(\omega) + 2i\beta^{res}(\omega)$ is much smaller than the nonresonant part $\chi^{nr} = -2\delta^{nr} + 2i\beta^{nr}$, the Fresnel formula for the reflectivity amplitude $R^{Fr}$

$$R^{Fr} = \frac{\sin\theta - \sqrt{\sin^2\theta + \chi}}{\sin\theta + \sqrt{\sin^2\theta + \chi}} \qquad (S2)$$

can be essentially simplified in the total reflection region, where $\text{Re}\,\chi^{nr}$ is large enough compared with $\sin^2\theta$ and negative. Let's insert the designation:

$$\sqrt{\sin^2\theta - 2\delta^{nr}} = i\rho , \qquad (S3)$$

and if $\beta = \beta^{nr} + \beta^{res} \ll \rho^2 = 2\delta^{nr} - \sin^2\theta$ and $\delta^{res} \ll \rho^2$, the approximate expression for the square roots in (S2) can be obtained:

$$\sqrt{\sin^2\theta + \chi} = \sqrt{\sin^2\theta - 2\delta^{nr} + 2i\beta^{nr} - 2\delta^{res} + 2i\beta^{res}} =$$
$$= \sqrt{\sin^2\theta - 2\delta^{nr}}\sqrt{1 + \frac{2i\beta - 2\delta^{res}}{\sin^2\theta - 2\delta^{nr}}} = \qquad (S4)$$
$$= i\rho\sqrt{1 + \frac{2i\beta - 2\delta^{res}}{-\rho^2}} \cong i\rho\left(1 + \frac{\delta^{res} - i\beta}{\rho^2}\right) = i\rho + i\delta^{res}/\rho + \beta/\rho$$

So (S2) takes a form:

$$R^{Fr} \cong \frac{\sin\theta - (i\rho + i\delta^{res}/\rho + \beta/\rho)}{\sin\theta + (i\rho + i\delta^{res}/\rho + \beta/\rho)} = \frac{\rho\sin\theta - \beta - i(\rho^2 + \delta^{res})}{\rho\sin\theta + \beta + i(\rho^2 + \delta^{res})} . \qquad (S5)$$

The squared module of (S5) gives the expression for the reflected intensity $\left|R^{Fr}\right|^2$:

$$\left|R^{Fr}\right|^2 \cong \frac{(\rho\sin\theta - \beta)^2 + (\rho^2 + \delta^{res})^2}{(\rho\sin\theta + \beta)^2 + (\rho^2 + \delta^{res})^2} \cong \frac{\rho^2(\sin^2\theta + \rho^2) - 2\rho\beta\sin\theta + 2\rho^2\delta^{res}}{\rho^2(\sin^2\theta + \rho^2) + 2\rho\beta\sin\theta + 2\rho^2\delta^{res}} =$$
$$= \frac{(\sin^2\theta + \rho^2) - 2\beta/\rho\,\sin\theta + 2\delta^{res}}{(\sin^2\theta + \rho^2) + 2\beta/\rho\,\sin\theta + 2\delta^{res}}$$

$$(S6)$$

where we neglect $\delta^{res2}$ and $\beta^2$. Taking into account that from (S3)

$$\sin^2\theta + \rho^2 = 2\delta^{nr}, \qquad (S7)$$

(S6) can be transformed to:

$$\left|R^{Fr}(\omega)\right|^2 \cong \frac{\delta^{nr} - \beta/\rho \sin\theta + \delta^{res}}{\delta^{nr} + \beta/\rho \sin\theta + \delta^{res}} \cong \left(1 - \frac{\beta \sin\theta}{\rho \delta^{nr}} + \frac{\delta^{res}}{\delta^{nr}}\right)\left(1 - \frac{\beta \sin\theta}{\rho \delta^{nr}} - \frac{\delta^{res}}{\delta^{nr}}\right) \cong$$

$$\cong \left(1 - \frac{2\sin\theta}{\rho \delta^{nr}}(\beta^{nr} + \beta^{res}(\omega))\right)$$

(S8)

That is the formula (5) in the article. It predicts the linear dependence of reflectivity in the total external reflection region on the imaginary part of the small resonant contribution to the refractive index.

2. *Reflectivity from a semi-infinite medium with a relatively high absorption in the presence of a small resonant scattering in the total external reflection region*

If $\beta^{nr} \gg \beta^{res}$ the expression (S6) is calculated by a more complicated way:

$$\left|R^{Fr}\right|^2 \cong \frac{(\rho \sin\theta - \beta)^2 + (\rho^2 + \delta^{res})^2}{(\rho \sin\theta + \beta)^2 + (\rho^2 + \delta^{res})^2} =$$

$$= \frac{(\sin\theta - \beta^{nr}/\rho - \beta^{res}/\rho)^2 + (\rho + \delta^{res}/\rho)^2}{(\sin\theta + \beta^{nr}/\rho + \beta^{res}/\rho)^2 + (\rho + \delta^{res}/\rho)^2} \cong$$

$$\cong \frac{(\sin\theta - \beta^{nr}/\rho)^2 + \rho^2 - 2\beta^{res}(\sin\theta - \beta^{nr}/\rho)/\rho + 2\delta^{res}}{(\sin\theta + \beta^{nr}/\rho)^2 + \rho^2 + 2\beta^{res}(\sin\theta + \beta^{nr}/\rho)/\rho + 2\delta^{res}} \cong$$

$$\cong \left(\frac{(\sin\theta - \beta^{nr}/\rho)^2 + \rho^2}{(\sin\theta + \beta^{nr}/\rho)^2 + \rho^2} + \frac{-2\beta^{res}(\sin\theta - \beta^{nr}/\rho)/\rho + 2\delta^{res}}{(\sin\theta + \beta^{nr}/\rho)^2 + \rho^2}\right) \times$$

$$\times \left(1 - \frac{2\beta^{res}(\sin\theta + \beta^{nr}/\rho)/\rho + 2\delta^{res}}{(\sin\theta + \beta^{el}/\rho)^2 + \rho^2}\right)$$

(S9)

Neglecting again $\delta^{res2}$ and $\beta^{res2}$, the nominator of the resonant term is calculated by the following way:

$$\left[ 2\delta^{res} - 2\beta^{res}(\sin\theta - \beta^{nr}/\rho)/\rho \right]\left[ (\sin\theta + \beta^{nr}/\rho)^2 + \rho^2 \right] -$$
$$- \left[ 2\delta^{res} + 2\beta^{res}(\sin\theta + \beta^{nr}/\rho)/\rho \right]\left[ (\sin\theta - \beta^{nr}/\rho)^2 + \rho^2 \right] =$$
$$= 2\delta^{res}\left[ (\sin\theta + \beta^{nr}/\rho)^2 + \rho^2 - (\sin\theta - \beta^{nr}/\rho)^2 - \rho^2 \right] -$$
$$- 2\beta^{res}(\sin\theta - \beta^{nr}/\rho)/\rho\left[ \sin^2\theta + 2\sin\theta\beta^{nr}/\rho + (\beta^{nr}/\rho)^2 + \rho^2 \right] -$$
$$- 2\beta^{res}(\sin\theta + \beta^{nr}/\rho)/\rho\left[ \sin^2\theta - 2\sin\theta\beta^{nr}/\rho + (\beta^{nr}/\rho)^2 + \rho^2 \right] =$$
$$= 8\delta^{res}\sin\theta\,\beta^{nr}/\rho - 4\beta^{res}\sin\theta/\rho\,(\sin^2\theta + \rho^2 + (\beta^{nr}/\rho)^2) +$$
$$+ 8\beta^{res}\beta^{nr}/\rho^2\sin\theta\beta^{nr}/\rho =$$
$$= 8\delta^{res}\sin\theta\,\beta^{nr}/\rho - 4\beta^{res}\sin\theta/\rho\,2\delta^{nr} - 4\beta^{res}\sin\theta/\rho(\beta^{nr}/\rho)^2 +$$
$$+ 8\beta^{res}(\beta^{nr}/\rho)^2\sin\theta/\rho =$$
$$= 8\delta^{res}\sin\theta\,\beta^{nr}/\rho - 4\beta^{res}\sin\theta/\rho\,(2\delta^{nr} - (\beta^{nr}/\rho)^2)$$

. (S10)

Finally, we get the formula (6) in the article:

$$\left|R^{Fr}(\omega)\right|^2 \cong \frac{(\sin\theta - \beta^{nr}/\rho)^2 + \rho^2}{(\sin\theta + \beta^{nr}/\rho)^2 + \rho^2} - \beta^{res}(\omega)\frac{4\sin\theta(2\delta^{nr} - (\beta^{nr}/\rho)^2)}{\rho\left[(\sin\theta + \beta^{nr}/\rho)^2 + \rho^2\right]^2} + \qquad (S11)$$
$$+ \delta^{res}(\omega)\frac{8\sin\theta\,\beta^{nr}}{\rho\left[(\sin\theta + \beta^{nr}/\rho)^2 + \rho^2\right]^2}$$

3. *Reflectivity at the exact critical angle in the presence of a small resonant scattering*

At the exact critical angle when

$$\sin^2\theta - 2\delta^{nr} = 0, \qquad (S12)$$

the Fresnel formula for the reflectivity amplitude is simplified to the expression:

$$R^{cr} = \frac{\sin\theta - \sqrt{\sin^2\theta + \chi}}{\sin\theta + \sqrt{\sin^2\theta + \chi}} = \frac{\sqrt{2\delta^{nr}} - \sqrt{i2\beta^{nr} + \chi^{res}}}{\sqrt{2\delta^{nr}} + \sqrt{i2\beta^{nr} + \chi^{res}}}\,. \qquad (S13)$$

If $\chi^{res}(\omega) = -2\delta^{res}(\omega) + 2i\beta^{res}(\omega)$ is small enough, taking into account that $\sqrt{2i} = (1+i)$, the square root in (S13) can be approximated by the expression

$$\sqrt{i2\beta^{nr}+\chi^{res}} \cong \sqrt{i2\beta^{nr}}\left(1+\frac{\chi^{res}}{i4\beta^{nr}}\right) = \sqrt{\beta^{nr}}(1+i)\left(1-i\frac{-2\delta^{res}+i2\beta^{res}}{4\beta^{nr}}\right) =$$
$$= \frac{1}{\sqrt{\beta^{nr}}}(1+i)\left(\beta^{nr}+\beta^{res}/2+i\delta^{res}/2\right) = \qquad (S14)$$
$$= \frac{1}{\sqrt{\beta^{nr}}}\left((\beta^{nr}+\beta^{res}/2-\delta^{res}/2)+i(\beta^{nr}+\beta^{res}/2+\delta^{res}/2)\right)$$

So, the reflectivity amplitude (S13) takes a form:

$$R^{cr} = \frac{\sqrt{2\delta^{nr}\beta^{nr}}-\left((\beta^{nr}+\beta^{res}/2-\delta^{res}/2)+i(\beta^{nr}+\beta^{res}/2+\delta^{res}/2)\right)}{\sqrt{2\delta^{nr}\beta^{nr}}+\left((\beta^{nr}+\beta^{res}/2-\delta^{res}/2)+i(\beta^{nr}+\beta^{res}/2+\delta^{res}/2)\right)}.$$
(S15)

Accordingly for the reflected intensity the following expression takes place:

$$\left|R^{cr}\right|^2 \cong \frac{\left(\sqrt{2\delta^{nr}\beta^{nr}}-(\beta^{nr}+\beta^{res}/2-\delta^{res}/2)\right)^2 + (\beta^{nr}+\beta^{res}/2+\delta^{res}/2)^2}{\left(\sqrt{2\delta^{nr}\beta^{nr}}+(\beta^{nr}+\beta^{res}/2-\delta^{res}/2)\right)^2 + (\beta^{nr}+\beta^{res}/2+\delta^{res}/2)^2}.$$
(S16)

The calculations of $(\beta^{nr}+\beta^{res}/2+\delta^{res}/2)^2$ in the first order of smallness of $\chi^{res}$ gives

$$(\beta^{nr}+\beta^{res}/2+\delta^{res}/2)^2 = (\beta^{nr}+\beta^{res}/2)^2 + (\delta^{res}/2)^2 + 2(\beta^{nr}+\beta^{res}/2)\delta^{res}/2 \cong$$
$$\cong \beta^{nr2}+\beta^{nr}\beta^{res}+\beta^{nr}\delta^{res}$$
(S17)

and correspondingly for $(\beta^{nr}+\beta^{res}/2-\delta^{res}/2)^2$ we have:

$$(\beta^{nr}+\beta^{res}/2-\delta^{res}/2)^2 \cong \beta^{nr2}+\beta^{nr}\beta^{res}-\beta^{nr}\delta^{res}. \qquad (S18)$$

So, their sum is simplified to:

$$(\beta^{nr}+\beta^{res}/2+\delta^{res}/2)^2 + (\beta^{nr}+\beta^{res}/2-\delta^{res}/2)^2 \cong 2\beta^{nr}(\beta^{nr}+\beta^{res}). \qquad (S19)$$

With this simplification and neglecting $\delta^{res2}$ and $\beta^{res2}$, (S16) can be presented in the form:

$$\left|R^{cr}\right|^2 \cong \frac{\delta^{nr} + \beta^{nr} + \beta^{res} - \sqrt{2\delta^{nr}\beta^{nr}}\left(1 + \beta^{res}/(2\beta^{nr}) - \delta^{res}/(2\beta^{nr})\right)}{\delta^{nr} + \beta^{nr} + \beta^{res} + \sqrt{2\delta^{nr}\beta^{nr}}\left(1 + \beta^{res}/(2\beta^{nr}) - \delta^{res}/(2\beta^{nr})\right)} \cong$$

$$\cong \left(\frac{\delta^{nr} + \beta^{nr} - \sqrt{2\delta^{nr}\beta^{nr}}}{\delta^{nr} + \beta^{nr} + \sqrt{2\delta^{nr}\beta^{nr}}} + \frac{\beta^{res} - \sqrt{2\delta^{nr}\beta^{nr}}\left(\beta^{res}/(2\beta^{nr}) - \delta^{res}/(2\beta^{nr})\right)}{\delta^{nr} + \beta^{nr} + \sqrt{2\delta^{nr}\beta^{nr}}}\right) \times$$

$$\times \left(1 - \frac{\beta^{res} + \sqrt{2\delta^{nr}\beta^{nr}}\left(\beta^{res}/(2\beta^{nr}) - \delta^{res}/(2\beta^{nr})\right)}{\delta^{nr} + \beta^{nr} + \sqrt{2\delta^{nr}\beta^{nr}}}\right)$$

(S20)

The nominator of the resonant term in the same order of smallness is calculated by the following way:

$$\left[\beta^{res}\left(1 - \sqrt{2\delta^{nr}\beta^{nr}}/(2\beta^{nr})\right) + \delta^{res}\sqrt{2\delta^{nr}\beta^{nr}}/(2\beta^{nr})\right]\left[\delta^{nr} + \beta^{nr} + \sqrt{2\delta^{nr}\beta^{nr}}\right] -$$

$$- \left[\beta^{res}\left(1 + \sqrt{2\delta^{nr}\beta^{nr}}/(2\beta^{nr})\right) + \delta^{res}\sqrt{2\delta^{nr}\beta^{nr}}/(2\beta^{nr})\right]\left[\delta^{nr} + \beta^{nr} - \sqrt{2\delta^{nr}\beta^{nr}}\right] \cong$$

$$\cong 2\beta^{res}\left(\sqrt{2\delta^{nr}\beta^{nr}} - \frac{\sqrt{2\delta^{nr}\beta^{nr}}}{2\beta^{nr}}(\delta^{nr} + \beta^{nr})\right) + 2\delta^{res}\frac{\sqrt{2\delta^{nr}\beta^{nr}}}{2\beta^{nr}}(\delta^{nr} + \beta^{nr}) =$$

$$= \sqrt{2\delta^{nr}\beta^{nr}}\left[\beta^{res}\left(1 - \frac{\delta^{nr}}{\beta^{nr}}\right) + \delta^{res}\left(1 + \frac{\delta^{nr}}{\beta^{nr}}\right)\right]$$

(S21)

Finally, the expression for the reflected intensity at the exact critical angle takes the form (formulas (8) and (9) in the article):

$$\left|R^{cr}(\omega)\right|^2 \cong$$

$$\cong \left|R^{nr}\right|^2 + \frac{\sqrt{2\delta^{nr}\beta^{nr}}}{(\delta^{nr} + \beta^{nr} + \sqrt{2\delta^{nr}\beta^{nr}})^2}\left[\delta^{res}(\omega)\left(1 + \delta^{nr}/\beta^{nr}\right) + \beta^{res}(\omega)\left(1 - \delta^{nr}/\beta^{nr}\right)\right].$$

(S22)

where the first term presents the reflectivity from a nonresonant media:

$$\left|R^{nr}\right|^2 = \frac{\delta^{nr} + \beta^{nr} - \sqrt{2\delta^{nr}\beta^{nr}}}{\delta^{nr} + \beta^{nr} + \sqrt{2\delta^{nr}\beta^{nr}}}.  \qquad (S23)$$

4. *Reflectivity from an ultrathin resonant layer*

The reflectivity amplitude from an ultrathin resonant layer placed under a reflecting medium can be presented as (M. A. Andreeva and B. Lindgren, JETP Letters **76**(12), 704(2002).):

$$R^{tot}(\omega) \cong R^s + E^2 r^d(\omega) \qquad (S24)$$

where $R^s$ is the reflectivity amplitude from an underlying substrate, $r^d(\omega)$ is the reflectivity amplitude of an ultrathin resonant layer, $E$ is the total electric field at the position of the resonant layer.

If $r^d(\omega)$ is small enough the expression for the reflected intensity $\left|R^{tot}(\omega)\right|^2$ from an ultrathin resonant layer above a substrate could be presented as:

$$\left|R^{tot}\right|^2 = (R^s + E^2 r^d)(R^{s*} + E^{*2} r^{d*}) \cong \left|R^s\right|^2 + R^s E^{*2} r^{d*} + R^{s*} E^2 r^d. \qquad (S25)$$

After separation of the real and imaginary parts in $R^s$, $E$ and $r^d$, i.e. presenting $R^s = R' + iR''$, $E = E' + iE''$ and $r^d = r' + ir''$, the calculations of the resonant addition to the total reflectivity give the following result:

$$\begin{aligned}
R^s E^{*2} r^{d*} + R^{s*} E^2 r^d &= \\
&= (R' + iR'')E^{*2}(r' - ir'') + (R' - iR'')E^2(r' + ir'') = \\
&= (R'r' + R''r'')(E^{*2} + E^2) + i(R''r' - R'r'')(E^{*2} - E^2) = \\
&= 2(R'r' + R''r'')(E'^2 - E''^2) + 4(E'E'')(R''r' - R'r'') = \\
&= \left(2R'(E'^2 - E''^2) + 4R''(E'E'')\right)r' + \left(2R''(E'^2 - E''^2) - 4R'(E'E'')\right)r''
\end{aligned} \qquad (S26)$$

Supposing the amplitude of the incident wave equals 1, and replacing $E$ in (S26) by $E = 1 + R^s = 1 + R' + iR''$ we get

$$\begin{aligned}
R^s E^{*2} r^{d*} + R^{s*} E^2 r^d &= \\
&= \left(2R'((1+R')^2 - R''^2) + 4R''^2(1+R')\right)r' + \left(2R''((1+R')^2 - R''^2) - 4R'R''(1+R')\right)r'' \\
&= 2\left(R'(1+R')^2 + R''^2(2+R')\right)r' - 2\left(R''(1 - R'^2 - R''^2)\right)r''
\end{aligned}$$
$$(S27)$$

Finally, the total reflectivity $\left|R^{tot}(\omega)\right|^2$ from an ultrathin resonant layer placed under a substrate takes a form (formula (15) in the article):

$$\left|R^{tot}(\omega)\right|^2 \cong \left|R^s\right|^2 + 2\left(R' + (2+R')\left|R^s\right|^2\right)r' + 2R''\left(1 - \left|R^s\right|^2\right)r'' \cong$$

$$\cong \left|R^s\right|^2 + Q_1 r'(\omega) + Q_2 r''(\omega)$$

(S28)

If $R^s$ in (S26), (S27), is the Fresnel reflectivity amplitude from substrate, for separation of its real and imaginary parts the square root in (S2) should be presented in the form:

$$\sqrt{\sin^2\theta + \chi^s} = \sin\theta\sqrt{1 - 2\delta^s/\sin^2\theta + 2i\beta^s/\sin^2\theta} =$$

$$= \sin\theta\left(\sqrt{\frac{\sqrt{u^2+v^2}+u}{2}} + i\sqrt{\frac{\sqrt{u^2+v^2}-u}{2}}\right) = \frac{\sin\theta}{\sqrt{2}}\left(\sqrt{q+u} + i\sqrt{q-u}\right),$$

(S29)

where $\delta^s$ and $\beta^s$ characterize the susceptibility of the substrate $\chi^s = -2\delta^s + 2i\beta^s$. In (S29) the following notations are used:

$$u = 1 - 2\delta^s/\sin^2\theta, \quad v = 2\beta^s/\sin^2\theta, \quad q = \sqrt{u^2+v^2}.$$

(S30)

With these notations the reflectivity amplitude from a substrate (S2) can be rewritten in the form:

$$R^s = \frac{\sin\theta - \sqrt{\sin^2\theta + \chi^s}}{\sin\theta + \sqrt{\sin^2\theta + \chi^s}} = \frac{\sqrt{2} - \left(\sqrt{q+u} + i\sqrt{q-u}\right)}{\sqrt{2} + \left(\sqrt{q+u} + i\sqrt{q-u}\right)} =$$

$$= \frac{\left(\sqrt{2} - \sqrt{q+u} - i\sqrt{q-u}\right)\left(\sqrt{2} + \sqrt{q+u} - i\sqrt{q-u}\right)}{\left(\sqrt{2} + \sqrt{q+u}\right)^2 + (q-u)}$$

$$= \frac{2 - q - u - q + u - i\sqrt{q-u}[(\sqrt{2} - \sqrt{q+u}) + (\sqrt{2} + \sqrt{q+u})]}{\left(\sqrt{2} + \sqrt{q+u}\right)^2 + (q-u)} =$$

$$= \frac{2(1 - q - i\sqrt{2}\sqrt{q-u})}{2(1 + q + \sqrt{2}\sqrt{q+u})} = \frac{1}{W}\left((1-q) - i\sqrt{2}\sqrt{q-u}\right)$$

(S31)

where we designate

$$W = 1 + q + \sqrt{2}\sqrt{q+u} \ . \tag{S32}$$

So, the real and imaginary parts of $R^s$ are determined by the expressions:

$$R' = \frac{(1-q)}{W}, \quad R'' = -\frac{\sqrt{2}\sqrt{q-u}}{W} \tag{S33}$$

Calculations of $Q_1$ with $R^s$ from (S31) give:

$$\begin{aligned} Q_1 &= 2\left[R'(1+R')^2 + R''^2(2+R')\right] = \\ &= \frac{2}{W^3}\left((1-q)(W+(1-q))^2 + \left(\sqrt{2}\sqrt{q-u}\right)^2(2W+(1-q))\right) = \\ &= \frac{2}{W^3}\left((1-q)(W+(1-q))^2 + \left(\sqrt{2}\sqrt{q-u}\right)^2(2W+(1-q))\right) = \\ &= \frac{2}{W^3}(1-q)\left[(W+1-q)^2 + 2(q-u)\right] + \frac{8}{W^2}(q-u) \end{aligned} \tag{S34}$$

After simplifying the expression in the parentheses:

$$\begin{aligned} &\left[(W+1-q)^2 + 2(q-u)\right] = \\ &\left[\left(1+q+\sqrt{2}\sqrt{q+u}+1-q\right)^2 + 2(q-u)\right] = \left(4 + 4\sqrt{2}\sqrt{q+u} + 2(q+u)\right) + 2(q-u) = \\ &= \left(4 + 4\sqrt{2}\sqrt{q+u} + 4q\right) = 4W \end{aligned} \tag{S35}$$

the expression for $Q_1$ (S34) is reduced to

$$Q_1 = \frac{8}{W^2}[(1-q)+(q-u)] = \frac{8(1-u)}{W^2} = \frac{16\delta^s}{W^2 \sin^2\theta} \ . \tag{S36}$$

Similarly, we calculate $Q_2$ coefficient with $R^s$ from (S31):

$$\begin{aligned} Q_2 &= 2R''(1 - R''^2 - R'^2) = -\frac{2\sqrt{2}}{W}\sqrt{q-u}\left(1 - \left(\frac{\sqrt{2}\sqrt{q-u}}{W}\right)^2 - \left(\frac{(1-q)}{W}\right)^2\right) = \\ &= -\frac{2\sqrt{2}}{W^3}\sqrt{q-u}\left(W^2 - 2(q-u) - (1-q)^2\right) \end{aligned} \tag{S37}$$

The expression in the parentheses can be essentially simplified:

$$\left(W^2 - 2(q-u) - (1-q)^2\right) = (1 + q + \sqrt{2}\sqrt{q+u})^2 + 2u - 1 - q^2 =$$
$$= 2(q+u) + 2q + 2\sqrt{2}\sqrt{q+u} + 2q\sqrt{2}\sqrt{q+u} + 2u =$$
$$= 4q + 4u + 2\sqrt{2}\sqrt{q+u}(1+q) = 2\sqrt{2}\sqrt{q+u}(\sqrt{2}\sqrt{q+u} + 1 + q) = 2W\sqrt{2}\sqrt{q+u}$$

(S38)

Taking into account that

$$\sqrt{q-u}\sqrt{q+u} = \sqrt{q^2 - u^2} = v, \tag{S39}$$

the expression for $Q_2$ (S37) gets the form:

$$Q_2 = -\frac{2\sqrt{2}}{W^3}\sqrt{q-u}\left(2W\sqrt{2}\sqrt{q+u}\right) = -\frac{8v}{W^2} = -\frac{16\beta^s}{W^2 \sin^2(\theta)} \tag{S40}$$

The expressions (S36), (S40) present the formulas (18), (19) in the article.